\documentclass[iop,apj,twocolappendix,numberedappendix]{emulateapj}

\usepackage{savesym}
\usepackage{amsmath}

\savesymbol{tablenum}

\usepackage{xspace}
\usepackage{xcolor}

\usepackage[multidot]{grffile}
\usepackage{rotating}
\usepackage{booktabs}
\usepackage{tabularx}
\usepackage{todonotes}
\usepackage{aas_macros}

\usepackage{tikz}

\usepackage[range-units=single,range-phrase= -- ,load-configurations=abbreviations]{siunitx}

\restoresymbol{SI}{tablenum}

\usepackage{amsmath}
\usepackage{mathtools}
\renewcommand{\la}{\ensuremath{\left\langle}}
\newcommand{\ra}{\ensuremath{\right\rangle}}

\newcommand{\lv}{\ensuremath{\left\lvert}}
\newcommand{\rv}{\ensuremath{\right\rvert}}

\newcommand{\lp}{\ensuremath{\left(}} 
\newcommand{\rp}{\ensuremath{\right)}}

\renewcommand{\vec}[1]{\boldsymbol{#1}}
\newcommand{\mat}[1]{\boldsymbol{\mathbf{#1}}}

\newcommand{\hconj}{\ensuremath{\dagger}}

\newcommand{\vx}{\vec{x}}
\newcommand{\vxt}{\tilde{\vec{x}}}

\newcommand{\vn}{\vec{n}}
\newcommand{\va}{\vec{a}}
\newcommand{\vv}{\vec{v}}
\newcommand{\vk}{\vec{k}}
\renewcommand{\vr}{\vec{r}} 

\newcommand{\mS}{\mat{S}}
\newcommand{\mN}{\mat{N}}
\newcommand{\mR}{\mat{R}}
\newcommand{\mC}{\mat{C}}
\newcommand{\mB}{\mat{B}}

\newcommand{\mF}{\mat{F}}
\newcommand{\mI}{\mat{I}}

\newcommand{\mQ}{\mat{Q}}
\newcommand{\mP}{\mat{P}}
\newcommand{\mQt}{\tilde{\mat{Q}}}

\newcommand{\mCt}{\tilde{\mat{C}}}

\newcommand{\mLambda}{\mat{\Lambda}}
\newcommand{\mLambdat}{\tilde{\mat{\Lambda}}}

\newcommand{\vnhat}{\hat{\vec{n}}}
\newcommand{\vuhat}{\hat{\vec{u}}}
\newcommand{\vzhat}{\hat{\vec{z}}}
\newcommand{\vu}{\vec{u}}

\newcommand{\brsc}[1]{{\ensuremath{\scriptscriptstyle \left(#1\right)}}}

\DeclareMathOperator{\sinc}{sinc}

\DeclareMathOperator{\tr}{tr}

\newcommand{\vis}{\ensuremath{V}}

\newcommand{\uvplane}{\ensuremath{uv}-plane\xspace}

\usepackage{hyperref}
\usepackage[capitalise,noabbrev,nameinlink]{cleveref}
\renewcommand{\eqref}[1]{\cref{#1}}

\newcommand{\tcm}{\SI{21}{\centi\metre}\xspace}
\DeclareSIUnit\jansky{Jy}

\newcommand{\KLfull}{Karhunen-Lo\`{e}ve\xspace}

\begin{document}

\sisetup{detect-all}
\title{All-Sky Interferometry with Spherical Harmonic Transit Telescopes}

\author{J. Richard Shaw$^{1,\dagger}$}
\author{Kris Sigurdson$^2$}
\author{Ue-Li Pen$^1$}
\author{Albert Stebbins$^3$}
\author{Michael Sitwell$^2$}
\affil{$^{1}$Canadian Institute for Theoretical Astrophysics, 60 St. George St., Toronto, ON M5S 3H8, Canada \\ 
        $^{2}$Department of Physics and Astronomy, University of British
  Columbia, Vancouver, BC V6T 1Z1, Canada \\
  $^{3}$Theoretical Astrophysics Group, Fermi National Accelerator Laboratory, Batavia, IL 60510, USA}
  
\email{$^\dagger$ jrs65@cita.utoronto.ca}  

\begin{abstract}

In this paper we describe the spherical harmonic transit telescope, a novel
formalism for the analysis of transit radio telescopes. This all-sky approach
bypasses the curved sky complications of traditional interferometry and so is
particularly well suited to the analysis of wide-field radio interferometers.
It enables compact and computationally efficient representations of the data
and its statistics that allow new ways of approaching important problems like
mapmaking and foreground removal. In particular, we show how it enables the
use of the \KLfull transform as a highly effective foreground filter,
suppressing realistic foreground residuals for our fiducial example by at least a factor twenty below the \tcm
signal 
even in highly contaminated regions of the sky.  This is despite the presence of
the angle-frequency mode mixing inherent in real-world instruments with
frequency-dependent beams. We show, using Fisher forecasting, that foreground
cleaning has little effect on 
power spectrum constraints compared to
hypothetical foreground-free measurements. Beyond providing a natural real-world
data analysis framework for \tcm telescopes now under construction and future experiments, this formalism allows accurate power spectrum
forecasts to be made that include the interplay of design constraints and
realistic experimental systematics with twenty-first century \tcm science.

\end{abstract}

\maketitle

\sisetup{detect-none}


\section{Introduction}
\label{sec:introduction}

Mapping the Universe with the \tcm line of neutral hydrogen will revolutionise
our view of the Universe. It holds the promise of unravelling the mysteries of
dark energy \citep{Chang2008,Wyithe2008}, unveiling the epoch of reionisation
(EoR) \citep{Furlanetto2006}, and perhaps even extending our view of the
cosmos out far enough to shine light on the primordial dark ages
\citep{2004PhRvL..92u1301L}. Rapidly probing large volumes of the Universe
requires new large wide-field telescopes along with powerful new digital
processing hardware such as
GMRT\footnote{\url{http://gmrt.ncra.tifr.res.in/}},
LOFAR\footnote{\url{http://www.lofar.org/}},
MWA\footnote{\url{http://www.mwatelescope.org/}},
Omniscope,
PAPER\footnote{\url{http://eor.berkeley.edu/}},
BAOBAB\footnote{\url{http://bao.berkeley.edu/}},
BAORadio\footnote{\url{http://groups.lal.in2p3.fr/bao21cm/}},
BINGO\footnote{\citet{BINGO}},
CHIME\footnote{\url{http://chime.phas.ubc.ca/}},
EMBRACE/EMMA\footnote{\citet{EMBRACE}} and
Tianlai\footnote{\url{http://tianlai.bao.ac.cn/}}.
In recent years it has become increasingly clear that
new methods of interpreting and analysing the data from these revolutionary
new instruments will be necessary to realize their scientific potential
\citep{Myers2003,Tegmark2009,Parsons2009,Liu2010,Liu2011,Parsons2012,Dillon2012}.

We describe here the spherical harmonic transit telescope, a new paradigm for
analysing wide-field transit telescopes in the spherical harmonic domain that
is naturally suited to mapping the \tcm Universe. Any telescope with fixed
pointing observes the sky transit through its field of view. The rotation
about the poles periodically over the course of a sidereal day creates a
linear correspondence between time $t$ and azimuthal angle $\phi$. We obtain a
simple mapping between the observed data and a linear combination of the
spherical harmonic coefficients $a_{lm}$ of the sky at fixed angular
wavenumber $m$, mediated by the angular response of each element. In what
follows, we elaborate and make precise this basic idea in the context of wide-
field interferometers, including the radial (frequency) direction.

This formalism diverges sharply with traditional characterizations of radio
interferometry that are better suited to observations with a narrow field of
view,  often assuming tracking of a particular source of interest, and
exploit the Fourier transform mapping between the sky and the \uvplane. What
we describe here is an all sky formalism for describing interferometry that
naturally incorporates the observable modes on the sky --- the spherical
harmonics.

The foremost challenge for any \tcm mapping experiment is separating the
cosmological signal from astrophysical contaminants which are $10^3$--$10^5$
times larger \citep{Furlanetto2006,Morales2009}. Conceptually this is simple
--- the primary foreground sources (diffuse synchrotron emission from the
Galaxy and emission from extragalactic point sources) are smooth as a function
of frequency, while the \tcm signal decorrelates quickly as each frequency
corresponds to a different radial slice of the Universe. To remove
foregrounds one just needs to model and remove the smooth frequency component
from their observations. Unfortunately, in practice, the large dynamic range
between the amplitude of the foregrounds and the \tcm signal makes several
real-world effects extremely problematic. While the properties of the
cosmological \tcm signal are thought to be well understood, the astrophysical
foregrounds are poorly constrained at the small angular and frequency scales
that will be probed by forthcoming \tcm experiments --- a successful technique
should be robust to uncertainties in foreground modelling. Of course, these
experiments will themselves help characterize the properties of real-world
foregrounds. More troublesome is the phenomena of angular-frequency mode
mixing: in any real experiment the shape of the beam on the sky will vary with
the observed frequency \citep{Liu2009}. This mode mixing makes simple
frequency only foreground removal methods ineffective in practice.  We show
below that the spherical harmonic transit telescope formalism can naturally
address the issues of model uncertainty and mode mixing, and enable efficient
and effective discrimination of the \tcm signal from obscuring foregrounds.

Any foreground removal method aims to find a subset of the data within which
there is significantly more \tcm signal than astrophysical foregrounds.
However, in the presence of mode-mixing, it is not obvious how to select
a basis which separates the two components --- what we would like is a method
which can automatically generate it. Just such a technique exists in the form
of the \KLfull (KL) transform. In this paper we show how the $m$-mode
formalism, described hence, makes the use of the KL transform  computationally
feasible.   The result is a remarkably effective and robust filter for
rejecting bright foregrounds and we demonstrate its effectiveness using
realistic simulations of the radio sky and a simple fiducial interferometer
configuration.

In Sec. \ref{sec:Formalism} we introduce the all-sky formalism that is the
basis for this technique. In Sec \ref{sec:Mapmaking} we discuss the map-making
process in the spherical harmonic transit telescope paradigm. In Sec.
\ref{sec:RepresentingStatistics} we discuss how to best represent statistics
of the cosmological \tcm signal and foregrounds in the measurement basis, and
in Sec. \ref{sec:ForegroundRemoval} we discuss how the \KLfull transform can
be used to detect faint signals in the presence of bright foregrounds.  In
Sec. \ref{sec:FisherAnalysis} we quantify the information lost due to
foreground removal using Fisher Analysis, and estimate errors on power
spectra.  We conclude in Sec. \ref{sec:Conclusion}.   In Appendix
\ref{app:models} we discuss the signal and foreground models we employ.  In
Appendix \ref{app:simulatedmaps} we describe how we create realistic
simulations of radio emission.


\section{Formalism}
\label{sec:Formalism}



In this section we introduce the $m$-mode formalism, a new description of the measurement process for transit interferometers.

In radio interferometry a visibility $V_{ij}$ is the instantaneous correlation
between two feeds $F_i$ and $F_j$. We will assume that we can take a linear
combination of the signal from a dual polarisation antenna, with no cross-
polarisation or polarisation leakage, such that we are sensitive only to the
total intensity (Stokes $I$) part of the sky. The fully polarised extension to
this work is also a tractable problem, we address this in a subsequent paper,
\cite{Shaw2013b}. At any instant, a visibility is given by
\begin{align}
\label{eq:nvis}
  \vis_{ij} & = \la F_i F_j^* \ra \notag \\
  & = \frac{1}{\Omega_{ij}}\int d^2\vnhat \, A_i(\vnhat) A_j^*(\vnhat)
  e^{2 \pi i \vnhat \cdot \vu_{ij}} T(\vnhat)
\end{align}
where $\vu_{ij} = (\vr_i - \vr_j) / \lambda$ is the spatial separation between
the two feeds divided by the observed wavelength (that is the separation in
the \uvplane), $\vnhat$ is the position on the celestial sphere, and
$A_i(\vnhat)$ gives the primary beam of feed $i$. In the above we have
normalised our visibilities such that they are temperature like, and we have
defined them in terms of the brightness temperature $T = \lambda^2 I / 2 k_b$
instead of the total intensity $I$. The quantity $\Omega_{ij} = \sqrt{\Omega_i
\Omega_j}$ is the geometric mean of the individual beam solid angles
\begin{equation}
\Omega_i = \int \lv A_i(\vnhat) \rv^2 d^2\vnhat
\end{equation}
which also gives the effective antenna area $A_\text{eff} \Omega = \lambda^2$.
This ensures that for a sky with uniform brightness temperature $T$ the auto-correlation
of an antenna $\vis_{ii} = T$ with our definition.

As the Earth turns both the primary beams and the baseline separations rotate
relative to the celestial sphere. This means the measured visibilities change
periodically with the sidereal day. We take this into account by explicitly
including the dependence on the azimuthal angle $\phi$ and by averaging over
each sidereal day.

The measured visibilities are also corrupted by instrumental noise for which
we add a noise term $n_{ij}(\phi)$. We assume the noise is stationary such
that its statistics are independent of $\phi$. Rewriting \cref{eq:nvis} in
terms of a transfer function $B_{ij}$ leaves the measured visibility as
\begin{equation}
  V_{ij}(\phi) = \int \! d^2\vnhat\, B_{ij}(\vnhat; \phi) T(\vnhat) + n_{ij}(\phi)
\end{equation}
where the transfer function is
\begin{equation}
  B_{ij}(\vnhat; \phi) = \frac{1}{\Omega_{ij}}A_i(\vnhat; \phi) A_j^*(\vnhat;
  \phi) e^{2 \pi i \vnhat \cdot\vu_{ij}(\phi)} \; .
\end{equation}

Taking advantage of the periodicity in $\phi$, we Fourier transform the system
\begin{align}
  V^{ij}_m &= \int \frac{d\phi}{2\pi} V_{ij}(\phi) e^{-i m \phi} \\
  & = \sum_{l m'}\int \frac{d\phi}{2\pi} B^{ij}_{l m'}(\phi) a_{l m'}
  e^{-i m \phi}+ n^{ij}_m
\end{align}
where to proceed to the second line we have inserted the spherical harmonic
expansions of both the sky, and the beam transfer function
\begin{align}
T(\vnhat) & = \sum_{lm} a_{lm} Y_{lm}(\vnhat) \; , \\
B_{ij}(\vnhat; \phi) & = \sum_{lm} B^{ij}_{lm}(\phi) Y_{lm}^*(\vnhat) \; .
\end{align}
Note that we have defined $B^{ij}_{lm}$ relative to the conjugate spherical
harmonic in order to simplify later notation. As the $\phi$ dependence simply
rotates the functions about the Earth's polar axis, the transfer function at
any $\phi$ is trivially $B^{ij}_{lm}(\phi) = B^{{ij}}_{lm}(\phi\!=\!0) e^{i m
\phi}$. Combined with the exponential factor in the integral, this simply
generates the Kronecker delta $\delta_{mm'}$, and we find
\begin{equation}
\label{eq:vis_unpol1}
V^{ij}_m = \sum_{l} B^{ij}_{l m} a_{l m}+ n^{ij}_m \; .
\end{equation}

This gives a simple description of how the observed sky maps into the measured
data given a telescope design (which is contained in the beam transfer
matrices $B^{ij}_{lm}$). This transformation does not mix $m$-modes on the
sky, and can therefore be performed on an $m$-by-$m$ basis --- for any
particular $m$ and frequency $\nu$ the measured visibilities are simply a
projection of the $l$-modes on the sky for the measured $m$. As the optical
system is of a finite size, this limits both the $l$ and $m$ to which the
telescope is sensitive, ensuring we only need to consider a finite number of
degrees of freedom, both measured ($V_m$) and on the sky ($a_{lm}$).

In fact whilst the positive and negative $m$-modes may be independent
measurements they are still observations of the same sky --- by transforming
the conjugate $V_{-m}^*$ and using that $a_{lm} = a^*_{l,-m}$ for a real field
we see that
\begin{equation}
V^{ij *}_{-m} = \sum_l (-1)^m B^{ij *}_{l, -m} a_{lm} + n^{ij *}_{-m} \; .
\end{equation}
In light of this we will change our notation such that we are considering only
the actual degrees of freedom on the sky. Let us separate out the positive and
negative $m$ parts by defining
\begin{align}
  B_{l m}^{ij, +} & = B^{ij}_{l m} & n^{ij, +}_{m} & = n^{ij}_{m} \\
  B_{l m}^{ij, -} & = (-1)^m B^{ij *}_{l, -m} & n^{ij, -}_m & = n^{ij *}_{-m}
\end{align}
which is valid for $m \ge 0$. Additionally to prevent double counting the $m =
0$ measurement we need to set $B_{l 0}^{ij -} = n_0^{ij -} = 0$. This gives a
modified version of \cref{eq:vis_unpol}
\begin{equation}
V_{m}^{ij,\pm} = \sum_l B_{l m}^{ij, \pm} a_{lm} + n_{m}^{ij, \pm} \; .
\end{equation}
For brevity of notation, we will introduce a label $\alpha$ which indexes both
the positive and negative $m$ parts of all included feed pairs $ij$, such that
any particular $\alpha$ specifies exactly the values of $ij,\pm$ (exactly how
$\alpha$ is packed is unimportant). This gives
\begin{equation}
\label{eq:vis_unpol}
V^\alpha_m = \sum_{l} B^\alpha_{l m} a_{l m}+ n^\alpha_m \; .
\end{equation}

The beam transfer matrices above can be written in an explicit matrix notation
\begin{equation}
\left(\mB_m\right)_{(\alpha \nu) (l \nu')} = B^{\alpha, \nu}_{lm} \delta_{\nu \nu'}
\end{equation}
where the row index labels all combinations of baseline ($\alpha$) and frequency
($\nu$), whereas the column index is over all multipole ($l$) and frequencies
($\nu'$). Similarly we can define vectors for the visibilities and harmonic coefficients
\begin{equation}
\left(\vv_m\right)_{(\alpha \nu)} = V^{\alpha, \nu}_m \, \quad
\left(\va_m\right)_{(l \nu)} = a_{lm}^\nu \; .
\end{equation}
From here onwards we'll drop the subscript $m$ denoting the
spherical harmonic order, all the equations below are valid for any $m$.
This allows us to rewrite \cref{eq:vis_unpol} as
\begin{equation}
\label{eq:matnot}
\vv = \mB\, \va + \vn \; .
\end{equation}

This simple linear description of the measurement process of a transit
telescope is extremely powerful. By reducing it down to a linear mapping
between a finite number of degrees of freedom it allows us to apply the
standard tools of signal processing. In the subsequent sections we apply it to
solve two challenging problems in \tcm radio astronomy.

\section{Map-making}
\label{sec:Mapmaking}

In astronomy being able to transform our measured signal into an accurate map
of the sky is essential. Whilst in this paper we explicitly avoid this process
for our analysis, preferring to carry it out directly in the data space, maps
are still needed for visualisation and cross-checking. Map-making with
interferometric data is generally a complicated process performed by
algorithms such as CLEAN \citep{CLEAN} and its derivatives. This is especially
true with wide fields of view where mosaicing and $w$-projection are generally
required. However, the $m$-mode formalism makes the map-making process on the
full sky conceptually simple.

First, we assume that the instrumental noise $\vn$ follows a complex gaussian
distribution with covariance $\mN = \la \vn \vn^\hconj \ra$, and the
different frequency channels are independent. For stationary noise, the
$m$-modes are uncorrelated and the likelihood function of the observed sky for
a single $m$ and frequency $\nu$ is
\begin{equation}
\label{eq:like1}
p(\vv \vert \va) = \frac{1}{\lv \pi \mN \rv} \exp{\lp - \lp\vv - \mB\va\rp^\hconj \mN^{-1} \lp \vv - \mB \va \rp \rp}
\end{equation}
where the vector $\va$ contains all harmonic coefficients for the given $m$.

To estimate the sky corresponding to a given set of visibilities we will look
for a maximum likelihood solution $d p / d \va = 0$. In particular we want
to find the value of $\va$ that minimises
\begin{equation}
\chi^2 = \lv \mN^{-\frac{1}{2}} \vv - \lp \mN^{-\frac{1}{2}} \mB \rp \va \rv^2 \; .
\end{equation}
The matrix $\mN^{-\frac{1}{2}}$ represents any factorisation such that
$(\mN^{-\frac{1}{2}})^\hconj \mN^{-\frac{1}{2}} = \mN^{-1}$. Provided $\mN$
contains no noiseless modes, it is positive-definite and so this factorisation
should exist. The maximum likelihood solution is given by the Moore-Penrose
pseudo-inverse \footnote{For details see \url{http://en.wikipedia.org/wiki/Moore-Penrose_pseudoinverse}}
\begin{equation}
\label{eq:mapmaking}
\hat{\va} = \lp \mN^{-\frac{1}{2}} \mB \rp^+ \mN^{-\frac{1}{2}} \vv \; ,
\end{equation}
where the superscript $+$ denotes the pseudo-inverse.

Depending on the number of baselines measured and the maximum $l$ we are
sensitive to, the problem may either be over- or under-constrained. In either
regime the Moore-Penrose pseudo-inverse gives a solution, in the former case
this reduces to the standard map making equation $\hat{\va} = (\mB^\hconj
\mN^{-1} \mB)^{-1} \mB^\hconj \mN^{-1} \vv$, and in the latter case selects
the solution which also minimises $\lv\hat{\va}\rv^2$, effectively setting
unconstrained degrees of freedom to zero.

As both distinct frequencies and $m$-modes are independent, map-making for a
set of full sky observations is a case of collating the estimates for each
individual $\nu$ and $m$.

\vspace*{10pt}


\section{Two Point Statistics}
\label{sec:RepresentingStatistics}

For Intensity Mapping experiments, our data has three components: the \tcm
signal which we are ultimately trying to extract, the foregrounds, and
instrumental noise.  Understanding the 2-point statistics of the data is of
paramount importance to our analysis --- not only do the correlations of the
signal encode most of the cosmological information that we are interested in
(see Appendix~\ref{app:models}), but to efficiently extract this we require
knowledge of the 2-point statistics of all three components. Here we write
down the linear relationship between these 2-point statistics of the data, and
how they are related to the underlying physical correlations.

The statistics of instrumental noise live in the visibility space, the basis
of our measurements. However the other components are naturally represented on
the sky, and must be projected into this space using \cref{eq:vis_unpol}. The
lowest non-zero moment of the visibilities is their covariance
\begin{multline}
\label{eq:signal_cv_full}
C_{(\alpha \nu m); (\alpha' \nu' m')} = \la V^m_{\alpha \nu} V^{m *}_{\alpha'
  \nu'} \ra \\
 = \sum_{l l'} B^{\alpha \nu}_{l m} \la a_{l m \nu}^{*} a_{l' m' \nu'} \ra
B^{\alpha' \nu' *}_{l' m'} + \la n_{(\alpha \nu m)} n^*_{(\alpha' \nu' m')} \ra \; .
\end{multline}
This is the covariance between all measured degrees of freedom: baselines,
frequencies, and $m$-modes. For the experiments listed in
Section~\ref{sec:introduction} we expect $\gtrsim 10^3$, $\sim 10^2$ and
$10^3$ respectively. This gives matrices of dimension $\gtrsim 10^8$, too
large to be tackled with current technology, both in terms of computation and
storage.

Instead, let us make an approximation that will dramatically reduce this
complexity. If we think of the sky as a statistically isotropic random field,
its two point statistics become dramatically simpler
\begin{equation}
\la a_{l m \nu'} a_{l' m' \nu'}^{*} \ra = C_l(\nu, \nu') \delta_{l l'}
\delta_{m m'} \; ,
\end{equation}
and importantly, they are automatically uncorrelated in the $m$ index. This
means that the full signal covariance \cref{eq:signal_cv_full} is block
diagonal and thus allows us to calculate all statistics on an $m$-by-$m$ basis. For a specific $m$-mode
\begin{equation}
C_{(\alpha \nu); (\alpha' \nu')} = \sum_{l} B^{\alpha \nu}_{l} B^{\alpha'
  \nu' *}_{l'} C_l(\nu, \nu') + N_{(\alpha \nu); (\alpha' \nu')} \; ,
\end{equation}
where we have dropped all the $m$-indices, and $N$ is the power spectrum on
the instrumental noise. As the number of $m$-modes we are sensitive to is
usually $\gtrsim 10^3$, assuming statistical isotropy saves at least a factor
of a million in computation and a thousand in storage. This ability to
efficiently perform calculations incorporating the full statistics opens up
new avenues for the data analysis of transit instruments. Synchrotron emission
from our galaxy clearly violates this assumption of statistical isotropy,
though, as we will demonstrate, this does not appear to affect our analysis and
in particular our ability to clean foregrounds.

In matrix notation
\begin{equation}
\mC = \mB \mC_\text{sky} \mB^\hconj + \mN \; .
\end{equation}
where we will split $\mC_\text{sky}$ into independent \tcm signal and
foreground parts $\mC_\text{sky} = \mC_{21} + \mC_f$.

The statistical models used for each component are chosen to be appropriate
for the frequency ranges of interest. In the fiducial example that follows
this is \SIrange{400}{600}{\mega\hertz}, corresponding to $z\sim 1$--$2$ for
the cosmological signal. These are described in Appendix~\ref{app:models}.

\section{Foreground Removal with the Karhunen-Lo\`{e}ve Transform}
\label{sec:ForegroundRemoval}

To clean our data we simply aim to find a subset within which there is
significantly more \tcm signal than the astrophysical foregrounds. However, in
the presence of mode mixing there is no immediately apparent representation in
which to perform this. This basis can be found using the \KLfull transform
(often called the Signal-to-Noise eigendecomposition), which has a long history
in Cosmology \citep[e.g.][]{Bond1994,Vogeley1996} and has been used for the analogous
problem of E/B mode separation for polarisation of the CMB
\citep{Lewis2002,Bunn2003}. This transform simultaneously diagonalises both
the signal and foreground covariance matrices, generating a set of modes with
no foreground or signal correlations. This makes comparing the amount of
signal and foreground power in each mode trivial.

To reduce the risk of foreground uncertainties biasing our analysis, we will
prioritise the removal of foreground contaminated modes at the expense of
cosmological signal. In contrast, the instrumental noise is well understood,
and we should be able to dig deeper into this contaminant to extract useful
cosmological information with little risk. In practice, this means we will
start with a filter which aggressively removes foregrounds only; subsequently
we will add back in the instrumental noise, which will allow us to compress
the data by removing completely noise-dominated modes, while retaining those
with a small fraction of signal.

This requires models for the statistics of both the signal and the
foregrounds. The signal is modelled as a simple gaussian random field for the
\tcm emission, whereas the foreground model includes both the synchrotron
emission from our galaxy, and the contribution from a background of
extragalactic point sources. The details of both are discussed in
Appendix~\ref{app:models}.

The \KLfull transform seeks to find a linear transformation of the data $\vv'
= \mP \vv$ such that the \tcm signal $\mS = \mB \mC_{21} \mB^\hconj$ and
foreground $\mF = \mB \mC_f \mB^\hconj$ covariance matrices are jointly
diagonalised. That is
\begin{equation}
\mS \rightarrow \mS' = \mP \mS \mP^\hconj = \mLambda \;,
\end{equation}
and
\begin{equation}
\mF \rightarrow \mF' = \mP \mF \mP^\hconj = \mI \; ,
\end{equation}
where $\mLambda$ is a diagonal matrix, and $\mI$ is the identity. In this
diagonal basis we can simply compare the amount of power expected in each mode
by the ratio of the diagonal elements (this is given by the corresponding
entries of $\mLambda$), and identify the regions of the space with low
foreground contamination (large entries in $\mLambda$).

This transformation can be found by solving the generalised eigenvalue problem
$\mS \vx = \lambda \mF \vx$. This gives a set of eigenvectors $\vx$, and
corresponding eigenvalues $\lambda$. Writing the eigenvectors in a matrix
$\mP$, row-wise, gives the transformation matrix to diagonalise the
covariances. The eigenvalues $\lambda$ corresponding to each eigenvector give
the diagonal matrix $\mLambda$

To isolate the \tcm signal, we want select modes with eigenvalue (signal-to-foreground
 power) greater than some threshold (see \cref{fig:evals}). To
project into this basis we define the matrix $\mP_s$ which contains only the
rows from $\mP$ corresponding to eigenvalues greater than the threshold $s$.

For most analysis we can work directly in the eigenbasis. However, for
visualising our results, we want to be able to transform back to the sky (by
way of the measured visibilities). To project back into the higher dimensional
space we simply generate the full inverse $\mP^{-1}$ and remove columns
corresponding to the rejected modes (we denote this matrix $\bar{\mP}_{s}$).
This is equivalent to projecting into the full eigenbasis, zeroing the
foreground contaminated modes, and then using the full-inverse $\mP^{-1}$.

For further analysis, we must include all noise terms, both foregrounds and
instrumental. Writing the total noise contribution as $\mN_\text{all} = \mF +
\mN$, the matrix in the truncated basis is
\begin{align}
\mN^\text{all} \rightarrow \mN^\text{all}_s & = \mP_s \lp \mF + \mN \rp \mP_s^\hconj \\
& = \mI + \mP_s \mN \mP_s^\hconj \; .
\end{align}
As the transformed instrumental noise matrix will not remain diagonal this
gives a correlated component between all our modes. However, as it is useful
if our modes are uncorrelated we make a further KL-transformation on the
foreground removed signal $\mS_s = \mLambda_s$, and total noise
$\mN^\text{all}_s$ covariance matrices. For computational and storage
efficiency we apply a further cut-off to include only modes with a signal to
total-noise ratio greater than cut-off value $t$. We denote this projection matrix $\mQt$.
For notational convenience we will write the total transformation in terms of
a single matrix $\mR = \mQ_t \mP_s$, having chosen suitable values for the two
cut-offs $s$ and $t$. As above, we will define an inverse $\bar{\mR} =
\bar{\mP}_s \bar{\mQ}_t$ which remains orthogonal to the removed space.
Quantities in this final basis we denote with tildes, for example a visibility
mapped into this basis is $\tilde{\vv} = \mR \vv$, and a covariance is $\mCt =
\mR \mC \mR^\hconj$.

\subsection{Cylinder Example}

While this method works with any transit telescope, to illustrate the
foreground removal process we will simulate a cylinder telescope, such as
CHIME or the Pittsburgh Cylinder Telescope \citep{PittsburghCylinders}. These
are transit interferometers composed of multiple parabolic cylinders where
each only focuses in the East-West direction. This gives a long and and thin
primary beam on the sky, extending nearly from horizon to horizon in the
North-South direction but which is only around 1 degree wide East-West.

Feeds are spaced along the axis of each cylinder --- when correlated these
provide resolution in the N-S direction. Correlations between cylinders
enhance the E-W resolution. The telescope operates as a transit telescope such
that the entire visible sky is observed once per sidereal day.

For a cylinder uniformly illuminated by a particular feed, near the axis the
beam pattern is a sinc function in the E-W direction, and uniform in the N-S
direction \cite[chapter 6]{RohlfsWilson}. To extend this off- axis we modulate
by projected area of the telescope giving
\begin{equation}
A^2(\vnhat) = \sinc^2{\lp \pi \, \vnhat\cdot\vuhat \frac{W}{\lambda} \rp} \, \Theta\lp\vnhat\cdot\vzhat\rp \, \vnhat\cdot\vzhat
\end{equation}
where $W$ is the cylinder width and $\vzhat$ is a unit vector pointing to the
zenith and $\vuhat$ is a unit vector pointing East in the ground-plane. The
step function $\Theta$ masks out the regions where the sky is below the
horizon, and the final factor $\vnhat\cdot\vzhat$ accounts for the projected
area of the telescope.

We model a two cylinder telescope observing the sky with 64 frequency channels
from \SIrange{400}{600}{\mega\hertz}. Each cylinder is \SI{15}{\metre} wide
and has 60 feeds regularly spaced by \SI{0.25}{\metre} (with the feeds lining
up E-W between cylinders). The telescope is located at a latitude of
\SI{45}{\degree}. These specifications correspond to a slightly smaller half-bandwidth version of the CHIME pathfinder telescope being
constructed at DRAO.

The noise covariance is diagonal for all $m$,
frequencies and baselines. For small $m$-modes with $m \ll 1 / \lp 2\pi \Delta\phi \rp$
(where $\Delta\phi$ is the angular integration time), the noise variance is
\begin{equation}
N^{ij}_m = \frac{T_{\text{sys},i}(\nu) T_{\text{sys},j}(\nu)}{4 \pi N_\text{day} t_\text{sid} \Delta\nu} \; ,
\end{equation}
where $T_{\text{sys},i}$ is the system temperature of a single polarisation of
feed $i$, $N_\text{day}$ is the number of sidereal days observed,
$t_\text{sid}$ is the length of a sidereal day, and $\Delta\nu$ is the
width of the frequency channel. As we combine the two polarisations into a
single unpolarised signal this reduces the noise power spectrum by a factor of
two. For this example $T_{\text{sys}} = \SI{50}{\kelvin}$, and we assume two
full years of observation (that is 730 complete sidereal days).

\begin{figure}

\includegraphics[width=\linewidth]{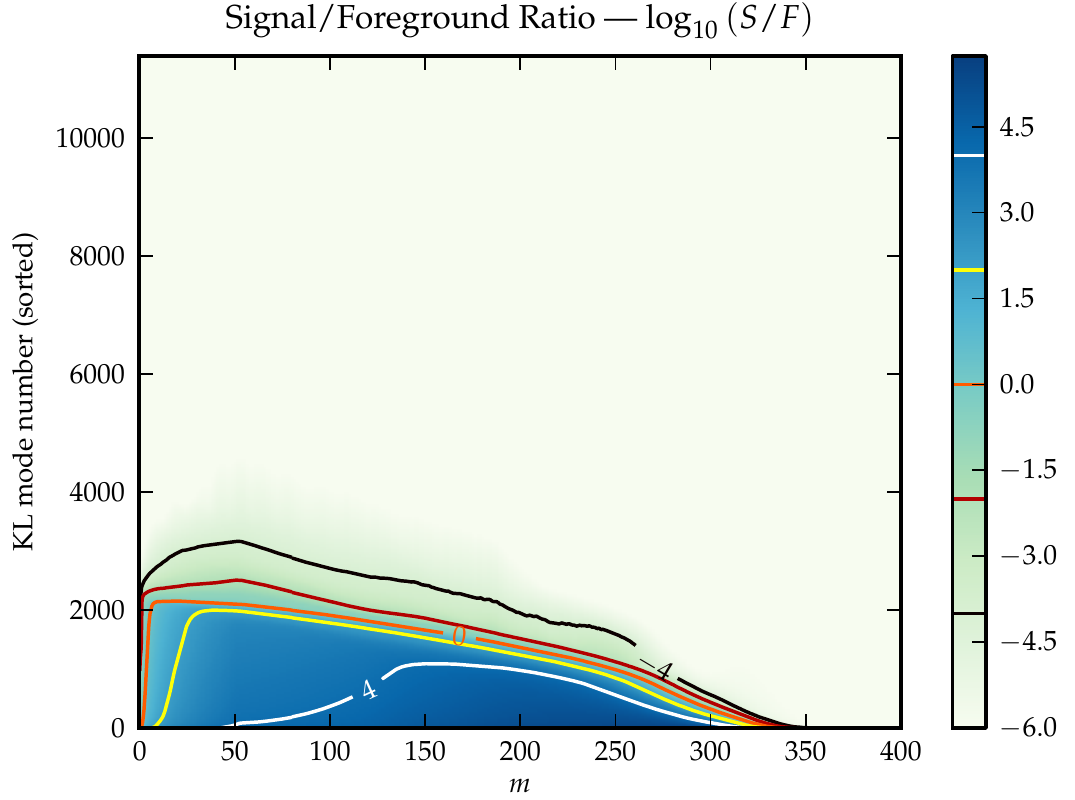}

\caption{The Signal-to-Foreground spectrum for all $m$-modes. We have 
plotted $\log_{10} \lambda_{i m}$, where for each $m$ the 
eigenvalues have been sorted by in ascending order (thus 
there is no physical interpretation to the vertical direction).
The contours are drawn at $-4$, $-2$, $0$, $2$ and $4$.}
\label{fig:evals}
\end{figure}

\newcommand{\tablabel}[1]{{\centering \textbf{\large #1}}}
\begin{figure*}[ht]
\begin{center}

\begin{tabularx}{\textwidth}{c@{\hspace{0.01\textwidth}}%
c@{\hspace{0.02\textwidth}}c@{\hspace{0.02\textwidth}}c}

&
\tablabel{Original} &
\tablabel{Observed} &
\tablabel{Foreground Filtered} \\

\begin{sideways}\hspace{2cm} \parbox{4.5cm}{ \tablabel{Foregrounds} }\hfill\end{sideways} &
\includegraphics[width=0.3\textwidth]{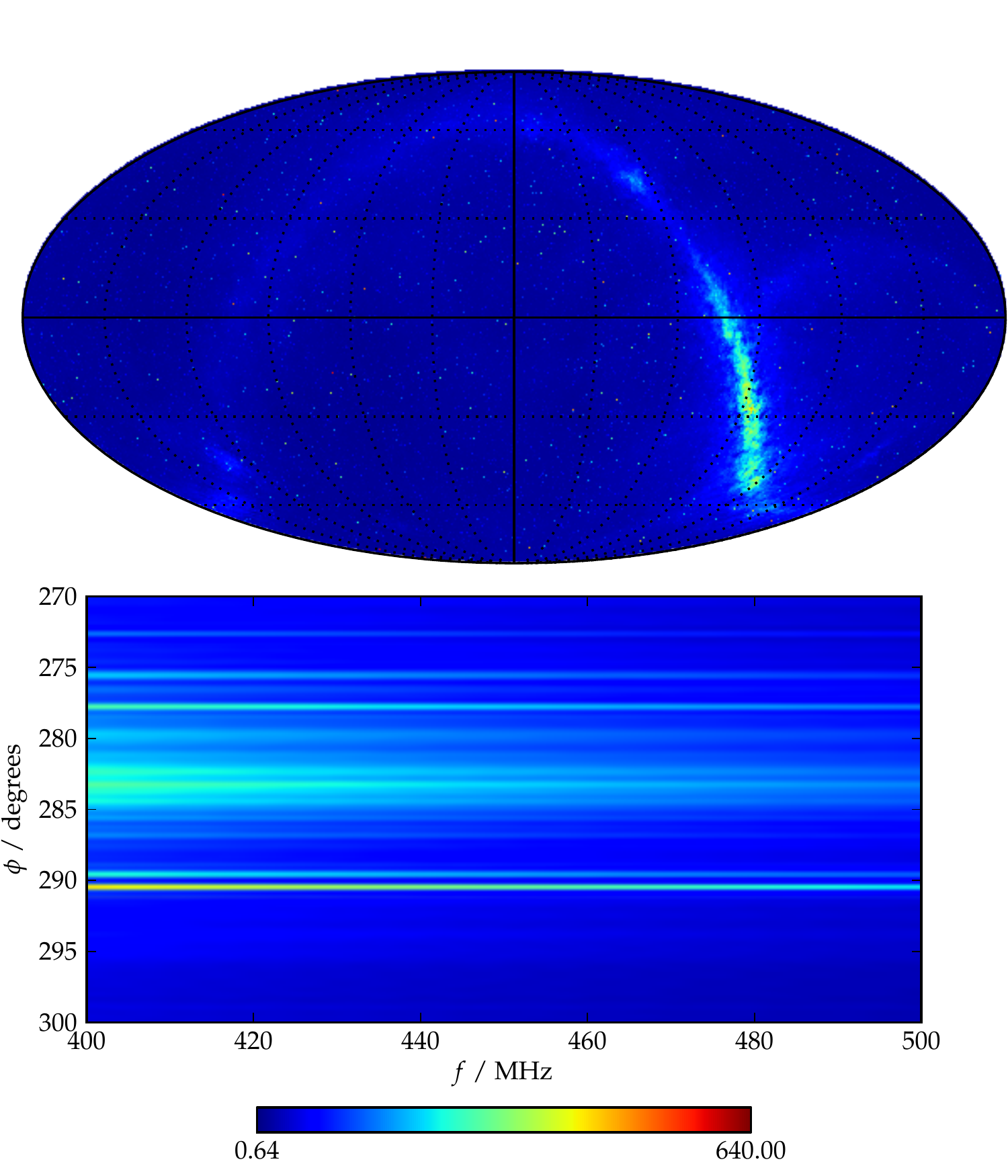} &
\includegraphics[width=0.3\textwidth]{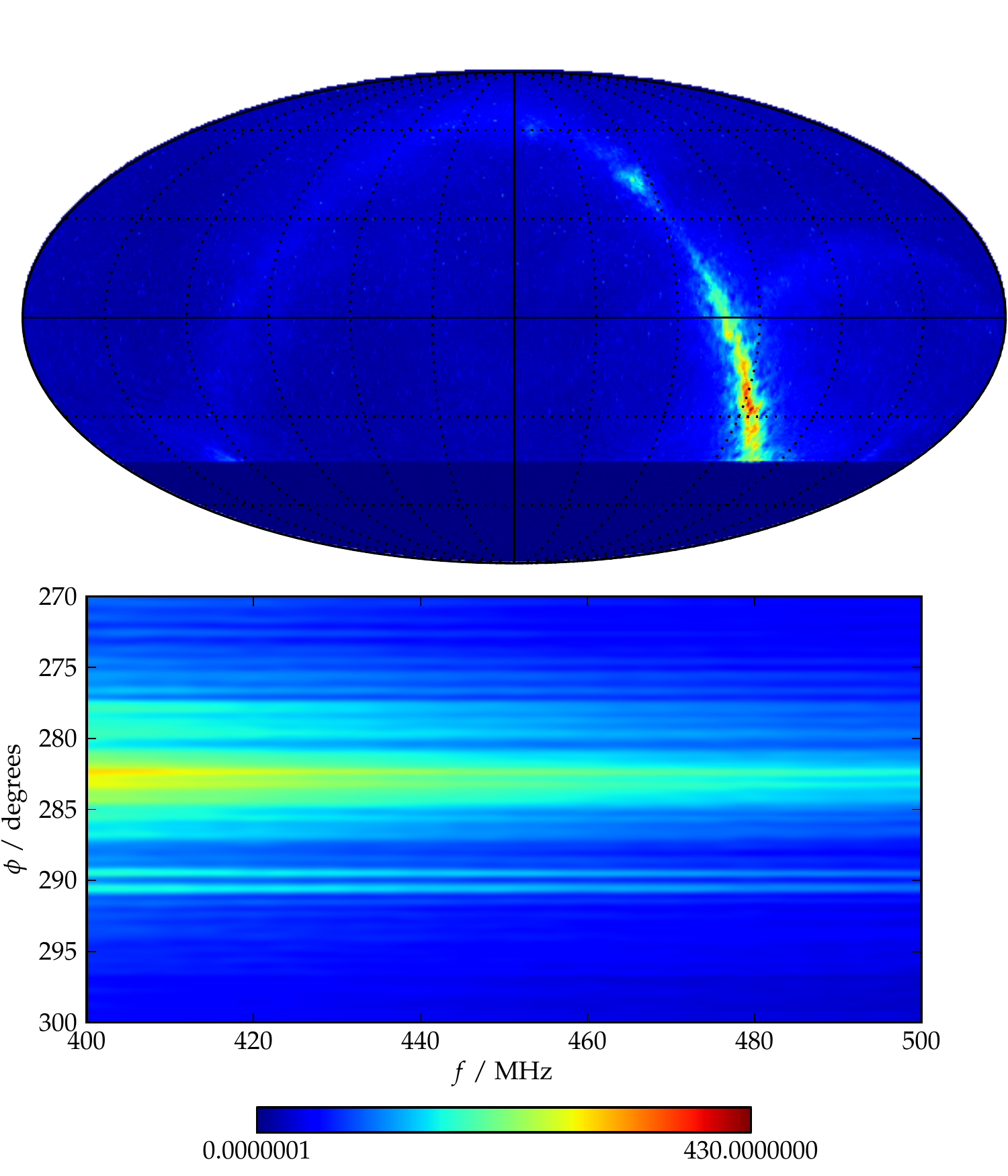} &
\includegraphics[width=0.3\textwidth]{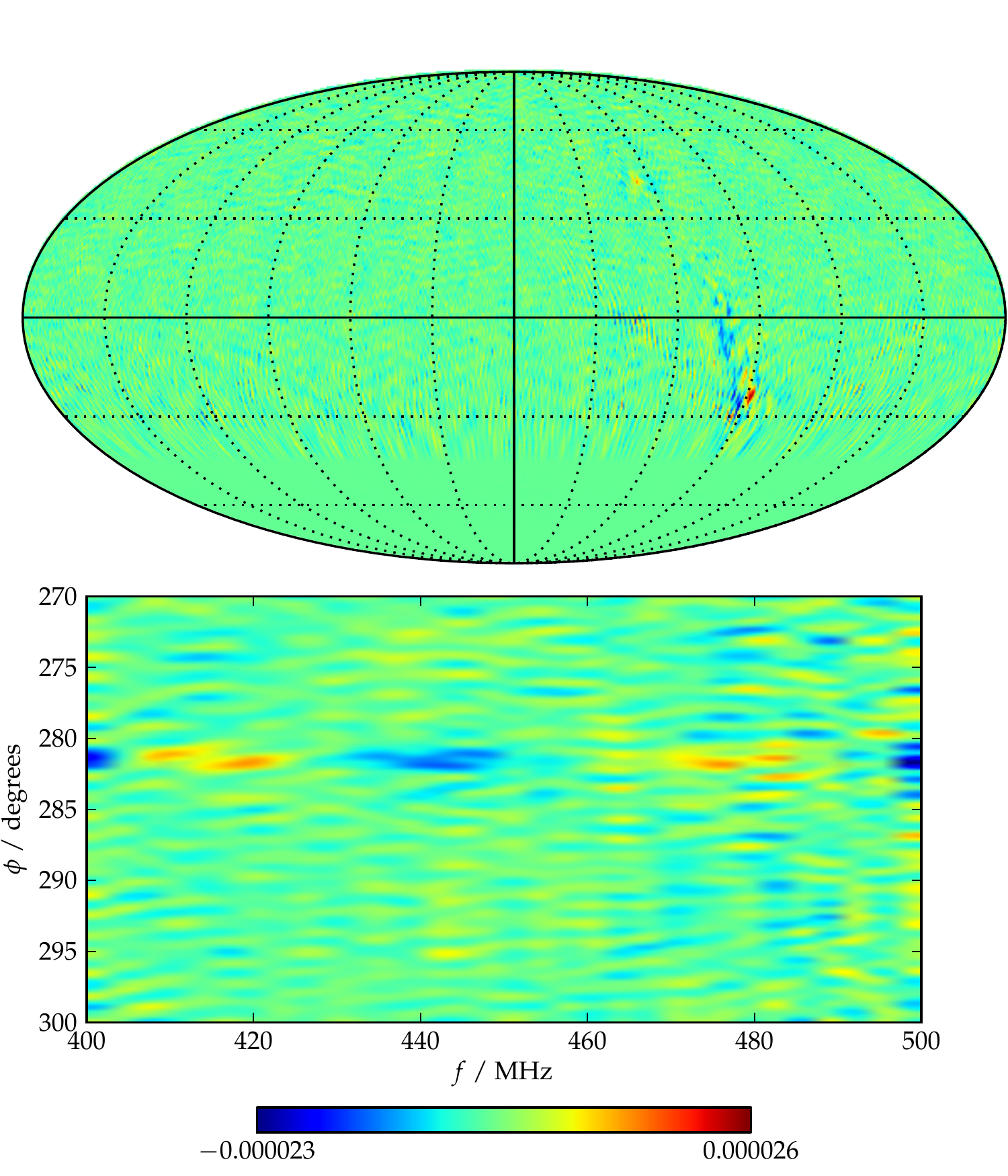} \\

\begin{sideways}\hspace{2cm} \parbox{4.5cm}{\tablabel{Signal}}\end{sideways} &
\includegraphics[width=0.3\textwidth]{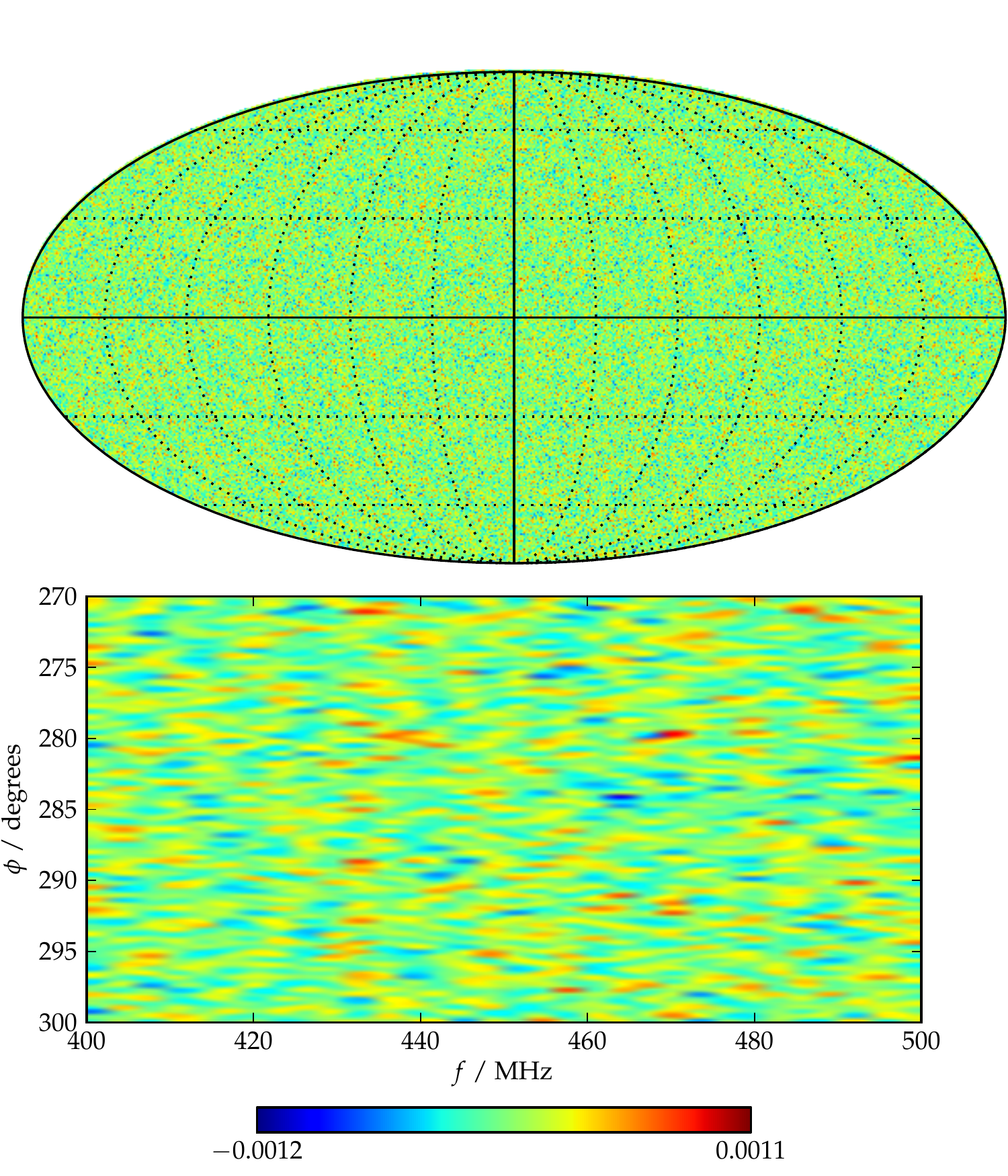} &
\includegraphics[width=0.3\textwidth]{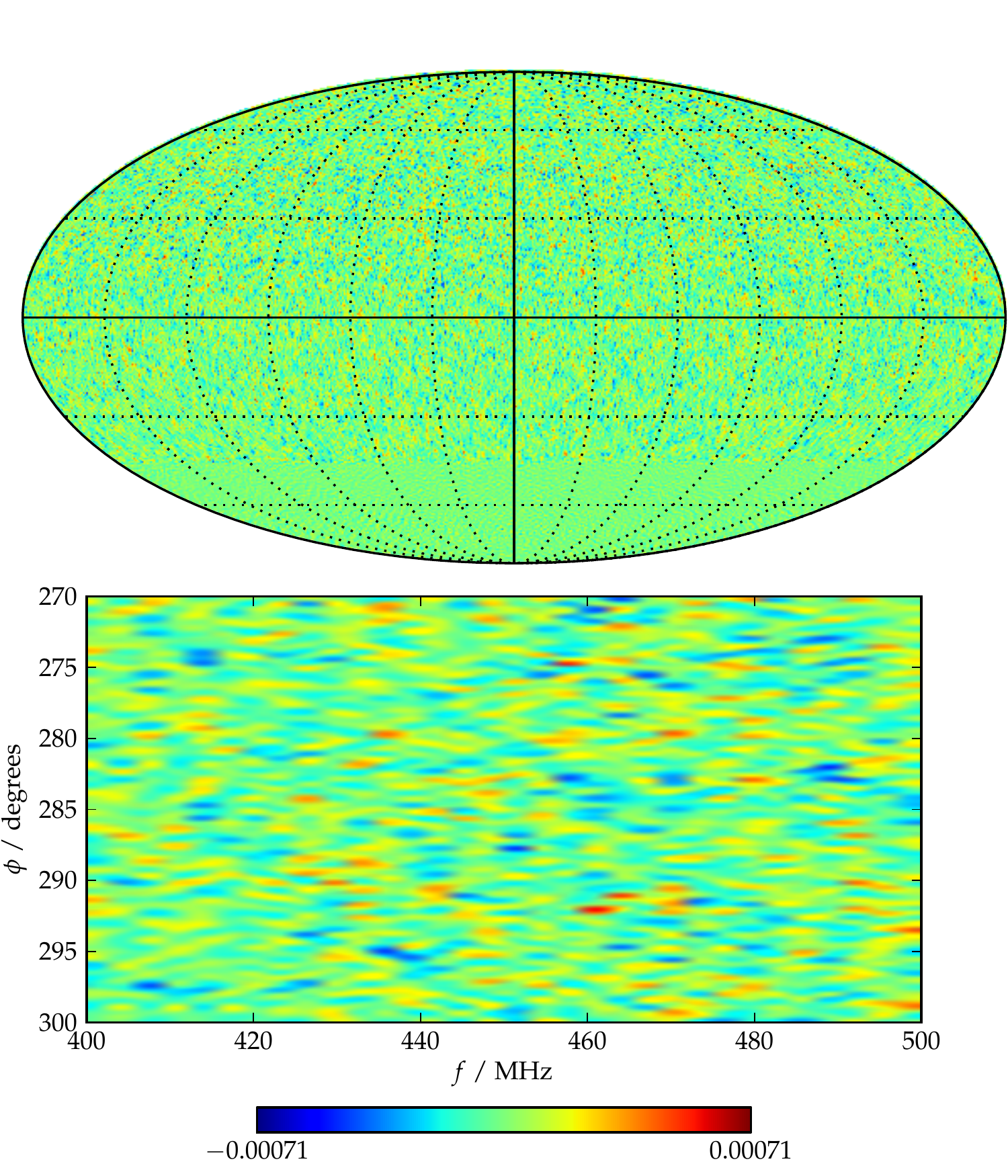} &
\includegraphics[width=0.3\textwidth]{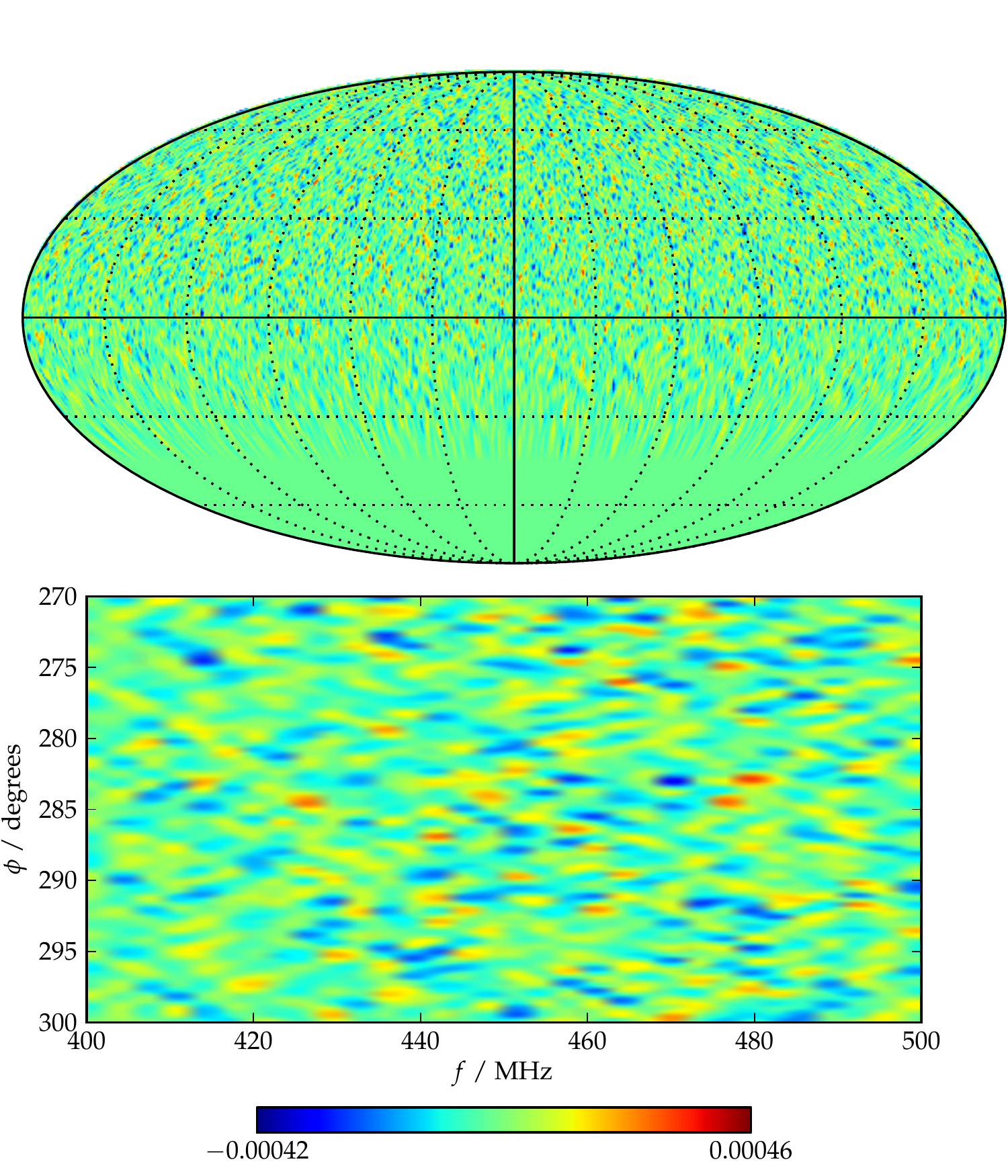}
\end{tabularx}
\end{center}

\caption{ This plot illustrates the foreground removal process in action on
simulations of the foregrounds-only (top row) and signal-only (bottom row).
Each plot has two elements, an image of the \SI{400}{\mega\hertz} frequency
slice on top, and beneath, a cut through the celestial equator (from
$270$--$300$ degrees) showing the frequency axis.  The left-most column shows
the original simulations on the sky. The band appearing in the foreground
frequency slice is the galactic plane. The middle column shows the maximum
likelihood map that we would make from the measured visibilities without
subtracting the low S/F modes. The maps are blank below
$\delta=\SI{-45}{\degree}$ because this area is always below the horizon for
the telescope at a latitude of \SI{45}{\degree}. The final column shows the
maps made after the foreground removal process (in this case we have discarded
modes with $S/F < 10$). This leaves a clear correspondence between the
original signal simulation and the foreground subtracted signal, whilst
leaving the foreground residuals over $20$ times smaller in amplitude. }

\label{fig:foregroundremoval}

\end{figure*}

In \cref{fig:evals} we show the spectrum of Signal-to-Foreground eigenvalues
for the telescope. The KL mode distribution of S/F has an extremely rapidly
rising spectrum so that the information retained (approximately the number of
modes) is rather insensitive to the cut threshold $s$ for values between
$10^{-2}$--$10^2$.

To demonstrate the foreground removal process we simulate time-streams from
separate realisations of the signal and foregrounds using \cref{eq:matnot},
and project them through the filtering process to make maps. The visibilities
are filtered using $\vv_\text{clean} = \bar{\mR} \mR \vv$, and then are turned
into a 3D map using \cref{eq:mapmaking}. In \cref{fig:foregroundremoval} we
show the original simulation, the map made from the unfiltered visibilities,
and the map made from the foreground filtered visibilities. The simulated
signal and foreground maps are described in Appendix~\ref{app:simulatedmaps}.
Note that the foreground maps are not simply realisations of the model used to
generate the foreground filter --- unlike the input model they are both non-
gaussian  and anisotropic. \Cref{fig:foregroundremoval} clearly illustrates
how the foreground amplitude is dramatically reduced by the process, whilst
the signal retains its overall character. Though the foreground residuals are
clearly highest in the galactic centre, even these are significantly lower
than the filtered signal.


\section{Fisher Analysis}
\label{sec:FisherAnalysis}

In the previous section we have demonstrated that the \KLfull transform gives
an effective method for removing foregrounds. Though a visual inspection of
\cref{fig:foregroundremoval} suggests that the \tcm signal is largely
untouched, we would like to be able to quantify how much useful information
remains. In this section we will use the Fisher matrix \citep[see][chap. 11
for an overview]{DodelsonBook} to forecast power spectrum errors, for the same
telescope, with and without foreground removal.

After projection into the reduced eigenbasis, let us assume that the remaining
modes follow a complex gaussian distribution with zero mean. This assumption
should be reasonable provided we have successfully removed the modes
containing any significant foreground contribution. In this case the Fisher
Information matrix for a set of parameters $p_a$ is
\begin{equation}
\label{eq:fisher_gaussian}
F_{ab}^\brsc{m} = \tr{\lp \mCt_a \mCt^{-1} \mCt_b \mCt^{-1} \rp} \; .
\end{equation}
where $\mCt_a = \partial \mCt / \partial p_a$. Though in the constructed
eigenbasis $\mCt = \mLambdat + \mI$ is diagonal, $\mCt_a$ can have
off-diagonal elements. Again this process is performed on a per-$m$ basis. As
there is no coupling between them, the total Fisher Information is simply the
sum over all $m$-modes
\begin{equation}
F_{ab} = \sum_m F_{ab}^\brsc{m} \; .
\end{equation}
For a set of parameters $p_a$ that we are trying to determine, the inverse of
the Fisher matrix is the lowest order approximation to their covariance.

\begin{figure}
\includegraphics[width=0.95\linewidth,page=1]{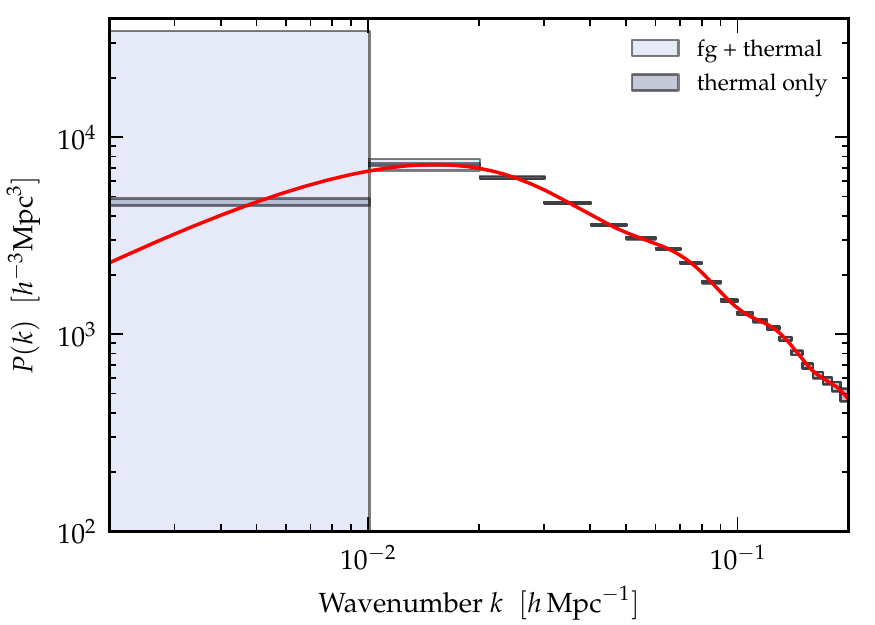}\\
\includegraphics[width=0.95\linewidth,page=2]{pkplots.pdf}

\caption{
\tcm Intensity Mapping provides a powerful technique for measuring the shape
of the matter power spectrum. In the plots above we illustrate the
power spectrum constraints that could be achieved with the large cylinder
telescope. The top plot shows the constraints on the whole power spectrum, the
lower plot zooms in on the region with the Baryon Acoustic Oscillations,
dividing through by a smoothed spectrum to remove the general trend. The dark
shaded bands are the errors we would find without foregrounds, where the only
noise is instrumental, the light bands include both contributions.
}

\label{fig:ps}

\end{figure}

In this work we will focus on forecasting the errors on the shape of the
matter power spectrum $P(k)$ whilst keeping all other cosmological parameters
fixed. Such forecasting has been performed using the \uvplane in \citet{Seo2010} and \citet{Ansari2012}.

We
parametrise the power spectrum in terms of a linear summation of different
basis functions
\begin{equation}
P(k) = \sum_a p_a P_a(k) \; .
\end{equation}
In Appendix~\ref{app:models} we describe how to project this quantity into the
angular power spectrum of \tcm fluctuations that we use to calculate the
visibility correlations. For simplicity, each of our bands is simply equal to
the input power spectrum within a fixed $k$-band, and zero outside, such that
the fiducial model is $p_a = 1$.

For the band-powers $p_a$ that we are trying to estimate, the matrices $\mCt_a$
are simply the projection of the basis functions $P_a(\vk)$ into the
eigenbasis. Starting from the angular power spectra
$C_{a; l}(\nu, \nu')$ corresponding to each of the basis functions $P_a(\vk)$
(using \eqref{eq:cl_flatsky})
\begin{equation}
\label{eq:ca_proj}
  \mCt_{a} = \mR \mB \mC_{21,a} \mB^\hconj \mR^\hconj \; .
\end{equation}
In practice explicitly calculating the $\mCt_a$ this way is computationally
expensive, we instead use a Monte-Carlo technique. We can form the estimator
$\hat{q}_a = \vxt^\hconj \mCt^{-1} \mCt_a \mCt^{-1} \vxt$, which has the
property that its covariance $\la \hat{q}_a \hat{q}_b \ra - \la \hat{q}_a \ra
\la \hat{q}_b \ra = F_{ab}$ \citep{Padmanabhan2003}. This means we can
estimate the $F_{ab}$ by averaging over realisations of $\vxt$. For
details see \cite{Dillon2012}.

In \cref{fig:ps} we plot the power spectrum errors for two cases: in the
presence of foregrounds that have been cleaned using our method and without
foregrounds at all. In the case without foregrounds, $\mF = 0$ and we only
perform the final \KLfull transform to diagonalise the signal and instrumental
noise. For the foregrounds we have cleaned modes with $S/F < 10$ and
additionally have removed modes with a small ratio of signal to total power.
This corresponds to setting $s=10$ and $t=0.01$. This is a clear demonstration
of the effectiveness of the technique --- it reduces our sensitivity on large
scales as we would expect (as the removed foreground are smooth on large
scales), while only slightly reducing our ability to constrain the small scale
power spectrum.


\section{Conclusion}
\label{sec:Conclusion}

In this paper we have introduced a powerful formalism for describing the
measurement process of transit telescopes (either interferometric or
otherwise). It is a natural formalism to describe interferometry on the full
sky --- sidestepping the standard complications that arise when dealing with
wide field interferometric data such as mosaicing and $w$-projection. A
spherical harmonic transit telescope allows for compact and computationally
efficient representations of the data and its statistics, which enable new
ways of approaching important problems like map-making and foreground removal.

Using the $m$-mode formalism and approximating the foregrounds as
statistically isotropic allows the powerful \KLfull transformation to be used,
automatically finding the basis in which the astrophysical foregrounds and
\tcm signal are maximally separated. The KL approach would be computationally
impossible otherwise and is a key advantage of the $m$-mode formalism. Using
this technique we can take the full three-dimensional dataset into account and
overcome the mode-mixing problem. The filters we construct are highly
effective and robust, a fact we have demonstrated by propagating through
realistically simulated \tcm and foreground timestreams. In our fiducial
example, shown in \cref{fig:foregroundremoval}, peak-to-peak foreground amplitude
was reduce by a factor of $\sim 2\times10^7$
leaving the peak-to-peak amplitude of the \tcm signal around $20$ times brighter that the foreground residuals.

We have also used this formalism to produce realistic forecasts for the power
spectrum constraints from a fiducial \tcm cylinder interferometer. We have
demonstrated that foreground cleaning does not significantly degrade \tcm
power spectrum estimates on BAO scales and below compared to a hypothetical
foreground-free measurement. We anticipate that the spherical harmonic transit
telescope formalism will be a powerful tool that can be applied to inform
experimental design and test the interplay between real-world systematics and
design constraints on twenty-first century \tcm science. We will explore this
further in \citet{Shaw2013b}.

\noindent 

\section*{Acknowledgements}

We thank the CHIME team for stimulating discussions, and Matt
Dobbs and Keith Vanderlinde for comments on an earlier version of this
manuscript. KS, UP, and MS are supported in part by the Natural Sciences and
Engineering Research Council (NSERC) of Canada. The work of AS was supported
by the DOE at Fermilab under Contract No. DE-AC02-07CH11359. KS thanks
Perimeter Institute for Theoretical Physics for their hospitality.
Computations were performed on the GPC supercomputer at the SciNet HPC
Consortium. SciNet is funded by: the Canada Foundation for Innovation under
the auspices of Compute Canada; the Government of Ontario; Ontario Research
Fund - Research Excellence; and the University of Toronto.

\appendix

\section{Signal and Foreground Models}

\label{app:models}

We model the \tcm signal and foregrounds as isotropic fields described by an
angular power spectrum $C_l(\nu, \nu')$. We base our models on
\cite{SantosCoorayKnox}, although we will only include the galactic
synchrotron and extragalactic point source contributions. Both these
contributions are assumed to take the form

\begin{multline}
\label{eq:aps_sck}
C_l(\nu, \nu') = A \lp \frac{l}{100}\rp^{-\alpha} \! \lp \frac{\nu \nu'}{\nu_0^2} \rp^{-\beta} \! e^{ - \frac{1}{2 \xi^2_l} \ln^2{(\nu / \nu')}} \, .
\end{multline}

As the models are calibrated for observations of the reionisation epoch, we
need to transform them into the higher frequencies we are concerned with. We
list the parameters for both these models in \cref{tab:modelparams}. 

For the point source model, which is based on the results of
\cite{DiMatteo2002}, we change the pivot frequency $\nu_0$ from \SI{150}{\MHz}
to \SI{408}{\MHz} and also rescale the amplitude in order to raise the maximum
flux of unsubtracted sources from \SI{0.1}{\milli\jansky} to
\SI{0.1}{\jansky}.

The galactic synchrotron model we use is not only calibrated for low
frequencies but also high galactic latitudes. As we will measure large
fractions of the sky we take this into account by changing the $A$ and angular
power-law index $\beta$ to be consistent with the angular power spectrum of
the \SI{408}{\MHz} Haslam map for galactic latitudes $\lv b \rv >
\SI{5}{\degree}$ from \cite{LaPorta2008}.

We model the 21cm brightness temperature as being a biased tracer of the
underlying matter fluctuations. These fluctuations are natually characterised
by the angular power spectrum \citep{Lewis2007,Datta2007}. However exact
calculation of this quantity requires double-integration over highly
oscillatory functions, instead we use the flat-sky approximation from
\cite{Datta2007}
\begin{equation}
\label{eq:cl_flatsky}
C_l(z, z') = \frac{1}{\pi \chi \chi'} \int_0^\infty \!\! dk_\parallel \cos{\lp k_\parallel \Delta\chi\rp} P_{T_b}(\vk; z, z')
\end{equation}
where $\chi$ and $\chi'$ are the comoving distances to redshift $z$ and $z'$.
Their difference is denoted by $\Delta\chi = \chi - \chi'$. The vector $\vk$
has the components $k_\parallel$ and $l / \bar{\chi}$ in the directions
parallel and perpendicular to the line of sight ($\bar{\chi}$ is the mean of
$\chi$ and $\chi'$). This approximation is accurate to the 1\% level for $l >
10$ \citep{Datta2007}.

We model the 21cm brightness temperature power spectrum $P_{T_b}$ as
\begin{equation}
P_{T_b}(\vk; z, z') = \bar{T}_b(z) \bar{T}_b(z') \lp b + f \mu^2 \rp^2 P_m(k; z, z')
\end{equation}
where $P_m(k; z, z') = P(k) D_+(z) D_+(z')$ is the real-space matter power
spectrum, $D_+(z)$ is the growth factor normalised such that $D_+(0) = 1$, $b$
is the bias, and the growth rate $f = d\ln{D_+} / d\ln{a}$, the logarithmic
derivative of the growth factor $D_+$. We assume that the bias $b=1$ at all
redshifts. The mean brightness temperature is assumed to take the form
\begin{multline}
\bar{T}_b(z) = 0.3 \lp \frac{\Omega_\text{HI}}{10^{-3}}\rp \\ \times\lp \frac{\Omega_m + (1+z)^{-3}\Omega_\Lambda}{0.29} \rp^{-1/2} \lp \frac{1+z}{2.5}\rp^{1/2} \si{\milli\kelvin}
\end{multline}
given in \cite{Chang2008}. We assume that the neutral hydrogen fraction takes
a value $\Omega_\text{HI} =\num{5e-4}$ \citep{Masui2013}.



\begin{table}

\caption{Our model for the angular power spectrum $C_l(\nu, \nu')$ is based on
those of \cite{SantosCoorayKnox} however we have adapted the parameters to
better suit the full-sky intensity mapping regime we are interested in.}

\label{tab:modelparams}
\begin{ruledtabular}
\begin{tabular}{l@{\hspace{0.04\linewidth}}l@{\hspace{0.04\linewidth}}l@{\hspace{0.04\linewidth}}l@{\hspace{0.04\linewidth}}l}
                & A (\si{\kelvin\squared})  & $\alpha$  & $\beta$   & $\zeta$   \\
\hline
Galaxy          & \num{6.6e-3}              & \num{2.80}& \num{2.8} & \num{4.0} \\
Point Sources   & \num{3.55e-4}             & \num{2.10}& \num{1.1} & \num{1.0} \\
\end{tabular}
\end{ruledtabular}

\end{table}


\section{Simulating All-sky Radio Emission}

\label{app:simulatedmaps}

\subsection{Galactic Synchrotron}

In order to test our methods we require simulated maps of the Galactic
emission from our own galaxy in the range \SIrange{400}{1400}{\mega\hertz}
with \SI{1}{\mega\hertz} resolution. Though there are maps at both
\SI{800}{\mega\hertz} and \SI{1420}{\mega\hertz}, the only public all-sky
radio survey in this range is the \SI{408}{\mega\hertz} Haslam map
\citep{HaslamMap}. However, the Global Sky Model \citep{GSM} is based on a
compilation of maps from \SI{10}{\mega\hertz} to \SI{94}{\giga\hertz}. We use
the Global Sky Model to generate maps at both \SI{400}{\mega\hertz} and \SI{1420}{\mega\hertz},
and use these to estimate an effective spectral at each location on the sky
\begin{equation}
\alpha(\vnhat) = \frac{\log{T_{1420}(\vnhat)} - \log{T_{400}(\vnhat)}}{\log{1420} - \log{400}} \; .
\end{equation}
By combining this with the Haslam map\footnote{We use the map from the Legacy Archive for Microwave Background Data Analysis (LAMBDA), which has
been processed to remove bright point sources and striping. See
\url{http://lambda.gsfc.nasa.gov/product/foreground/haslam_408.cfm}} we can
extrapolate to simulate a map of the sky at any desired set of frequencies
\begin{equation}
T_\text{base}(\vnhat, \nu) = T_{408}(\vnhat) \left(\frac{\nu}{\SI{408}{\mega\hertz}}\right)^{\alpha(\vnhat)} \; .
\end{equation}
Unfortunately this map lacks both the small scale angular fluctuations
(because of the limited resolution of the Haslam map) and any spectral
variations (because of the power law extrapolation) that would be present on
the real sky. It is essential to include these to accurately test any
foreground removal method.

To include these fluctuations we could add gaussian realisations of
\eqref{eq:aps_sck} (with the galactic synchrotron parameters, see
\cref{tab:modelparams}) to the base map, which contain frequency and angular
fluctuations at arbitrary resolutions. However the Haslam map already
constrains what the sky looks like on scales $\gtrsim \SI{1}{\degree}$, and
the extrapolation with the spectral index map, is a
constraint on the sky at \SI{1420}{\mega\hertz} (on scales larger than
\SI{5.1}{\degree}, the resolution of the Global Sky Model). Therefore we would
like the combined simulated map to be consistent with these observations. We
can do this by constraining the realisations to ensure there are no
fluctuations on the scales constrained. In practice we do this by manipulating
the amplitudes of the two highest valued eigenmodes of $C_l(\nu, \nu')$
(from Equation~\ref{eq:aps_sck}) in each realisation, to ensure that the \SI{408}{\MHz} and
\SI{1420}{\MHz} slices are zero when smoothed on \SI{1}{\degree} and
\SI{5.1}{\degree} scales respectively.

A further problem is that we know the amplitude of small scale fluctuations
varies over the sky, however our realisations are statistically
isotropic. This is clearly demonstrated in the analysis of \cite{LaPorta2008},
which shows that the amplitude of the angular power spectrum traces the
galatic structure. To reproduce this we use the RMS amplitude of
fluctuations across the Haslam map in $\sim \SI{4}{\degree}$ patches
(corresponding to Healpix pixels with $\mathrm{NSIDE} = 16$), to rescale the
fluctuations.  In particular this generates an angular power spectrum on the
whole sky which is consistent with a single power-law even when crossing
through the beam-scale of the Haslam map into the simulated fluctuations. We do
not include variations of the power-law index of the angular power spectrum as
there appears to be no structure to the small variations found in
\cite{LaPorta2008}.

\subsection{Extra-Galactic Point Sources}

Our point source maps are constructed from two components, a population of
bright  point sources ($S > \SI{0.1}{\jansky}$ at \SI{151}{\mega\hertz})
simulated directly, and a background of dimmer unresolved point sources ($S <
\SI{0.1}{\jansky}$) modelled as a gaussian random field.

The former is constructed directly by drawing from the point source
distribution of \cite{DiMatteo2002}, each sourced is modelled as having pure
power law emission with a random spectral index. The sources are distributed
randomly over the sky. Very bright sources ($S > \SI{10}{\jansky}$) are
assumed to have been subtracted such that their residuals are less than this
threshold.

The unresolved background is simulated by drawing a gaussian realisation from
\cref{eq:aps_sck} with the point source model detailed in
\cref{tab:modelparams}.

\subsection{\tcm Signal}

Simulations of the Cosmological 21cm emission are performed by drawing
gaussian realisations from the flat-sky angular power spectrum (calculated
using \cref{eq:cl_flatsky}).


\bibliographystyle{yahapj}
\bibliography{cyl}

\end{document}